\documentclass[journal]{IEEEtran}

\usepackage[utf8]{inputenc}
\usepackage[english]{babel}
\usepackage{dblfloatfix} 
\usepackage{amsmath, amsfonts, amssymb, amsthm}
\usepackage{booktabs}
\usepackage{enumitem}

\newcommand{\svgl}{$\textrm{svGL}$}
\newcommand{\mvgl}{$\textrm{mvGL}$}
\newcommand{\mvglone}{$\textrm{mvGL}_{\ell_1}$}
\newcommand{\mvgltwo}{$\textrm{mvGL}_{\ell_2}$}

\usepackage{multidef}
\usepackage{mathtools}

\multidef[prefix=set]{\ensuremath{\mathbb{#1}}}{A-Z}

\multidef[prefix=cal]{\ensuremath{\mathcal{#1}}}{A-Z}

\multidef[prefix=v]{\ensuremath{\mathbf{#1}}}{a-z}
\newcommand{\vell}{\boldsymbol{\ell}}

\multidef[prefix=m]{\ensuremath{\mathbf{#1}}}{A-Z}
\newcommand{\mLambda}{\boldsymbol{\Lambda}}

\newcommand{\ones}{\boldsymbol{1}}
\newcommand{\zeros}{\boldsymbol{0}}
\newcommand{\upper}{\mathrm{upper}}
\newcommand{\diag}{\mathrm{diag}}
\newcommand{\trace}{\mathrm{tr}}
\DeclareMathOperator*{\minimize}{minimize}
\DeclarePairedDelimiter\norm{\lVert}{\rVert}

\DeclareMathOperator*{\argmin}{argmin}

\newcommand{\ermodel}{Erdős–Rényi}
\newcommand{\bamodel}{Barabási–Albert}

\usepackage{enumitem}
\usepackage{pifont}
\newlist{todolist}{itemize}{2}
\setlist[todolist]{label=$\square$}

\usepackage{xcolor}

\newcommand{\currval}[1]{\Hat{#1}}
\newcommand{\prevval}[1]{\Hat{\Hat{#1}}}

\begin{document}

\title{Multiview Graph Learning with Consensus Graph}

\author{Abdullah Karaaslanli, Selin Aviyente,~\IEEEmembership{Senior Member,~IEEE}
\thanks{This work was supported in part by National Science Foundation under Grants CCF-2211645.}
\thanks{The authors are with the Department of Electrical and Computer Engineering, Michigan State University, East Lansing, MI 48824 USA (e-mail: karaasl1@msu.edu; aviyente@egr.msu.edu).}
\thanks{This work has been submitted to the IEEE for possible publication. Copyright may be transferred without notice, after which this version may no longer be accessible.}}

\markboth{}%
{Shell \MakeLowercase{\textit{et al.}}: A Sample Article Using IEEEtran.cls for IEEE Journals}


\maketitle
\begin{abstract}
Graph topology inference, i.e., learning graphs from a given set of nodal observations, is a significant task in many application domains. Existing approaches are mostly limited to learning a single graph assuming that the observed data is homogeneous. This is problematic because many modern datasets are heterogeneous or mixed and involve multiple related graphs, i.e., multiview graphs. Recent work proposing to learn multiview graphs ensures the similarity of learned view graphs through pairwise regularization, where each pair of views is encouraged to have similar structures. However, this approach cannot infer the shared structure across views. In this work, we propose an alternative method based on consensus regularization, where views are ensured to be similar through a learned consensus graph representing the common structure of the views. In particular, we propose an optimization problem, where graph data is assumed to be smooth over the multiview graph and the topology of the individual views and that of the consensus graph are learned, simultaneously. Our optimization problem is designed to be general in the sense that different regularization functions can be used depending on what the shared structure across views is. Moreover, we propose two regularization functions that extend fused and group graphical lasso to consensus based regularization. Proposed multiview graph learning is evaluated on simulated data and shown to have better performance than existing methods. It is also employed to infer the functional brain connectivity networks of multiple subjects from their electroencephalogram (EEG) recordings. The proposed method reveals the structure shared by subjects as well as the characteristics unique to each subject.
\end{abstract}
\begin{IEEEkeywords}
Multiview Graphs, Graph Inference, Graph Signal Processing
\end{IEEEkeywords}
\section{Introduction}
\label{sec:introduction}
Many real-world data are represented through the relations between data samples or features, i.e., a graph structure \cite{barabasi2002new,newman2018networks}. Although many datasets, including social and traffic networks, come with a known graph that helps in their interpretation, there is still a large number of applications where a graph is not readily available. For instance, functional or structural brain networks \cite{sporns2012discovering} and regulatory networks of genes \cite{li2006gradient} are not directly observable. In such cases, inferring the topology of the graph is an essential task to be able to effectively analyze their data.

In order to address this problem, various \textit{graph learning} techniques have been developed by using observed nodal data, also known as \textit{graph signals} \cite{ortega2018graph}, to learn the unknown graph topology. These methods are developed using approaches from different domains including statistical methods \cite{drton2017structure}, graph signal processing (GSP) \cite{dong2019learning, mateos2019connecting} and recently graph neural networks \cite{kipf2018neural}. Statistical methods, such as graphical lasso \cite{friedman2008sparse}, are usually based on Gaussian graphical models, where the aim is to learn the precision matrix representing conditional dependencies between nodes. Methodologies using GSP defines the relation between graph signals and graph topology using signal processing concepts, such as stationarity \cite{thanou2017learning, segarra2017network, pasdeloup2017characterization}, non-stationarity \cite{shafipour2021identifying} and smoothness \cite{dong2016learning, kalofolias2016learn, berger2020efficient}. Finally, GNN based techniques use recent graph convolutional networks and the more general message passing networks for relational inference \cite{kipf2018neural, franceschi2019learning, chen2020iterative}.

Although aforementioned methods have shown to be effective, they are limited to homogeneous datasets, where observed graph signals are assumed to be identically distributed and defined on a single graph. However, in many applications, the data may be heterogeneous or mixed and come from multiple related graphs, also known as \textit{multiview graphs}. For example, gene expression measurements are often collected across different cell types each with their own gene regulatory mechanism \cite{zhang2023inference, karaaslanli2023kernelized}. Similarly, neuroimaging data from multiple subjects can be considered as being defined on a multiview graph where each view is a subject's brain connectome \cite{betzel2019community, de2017multilayer}. In these situations, the views of the multiview graphs are closely related to each other. Therefore, learning the topology of views jointly by incorporating the relationships among views can improve the performance \cite{tsai2022joint, danaher2014joint, navarro2022joint}. 

In this setting of joint learning, two important questions arise. The first question is regarding how to quantify the similarity across the different views. The choice of the optimal similarity function depends on how views are related to each other and what the structures that are shared across views and unique to each view are. Prior work answers this question by assuming the similarity of topologies \cite{danaher2014joint, navarro2022joint, guo2011joint}, nodes' connectivities \cite{mohan2014node}, or communities \cite{arroyo2021inference} \textit{etc}. The second question is related to how to incorporate the similarity assumption into the learning algorithm. Existing work mostly uses pairwise regularization, where pairs of views are penalized to be similar to each other. Although this approach can incorporate similarity of views into the learning algorithm, it cannot reveal the shared structure across views. This may be important in certain applications such as network neuroscience, where both the subject level connectomes as well as a group connectome that summarizes what is common across subjects for a given task are essential to identify individual variation \cite{li2023computing}.

In this paper, we introduce a flexible multiview graph learning approach \textit{(i)} to allow for modeling different types of view similarities, and \textit{(ii)} to incorporate the similarity of views into the learning algorithm while learning the (un-)shared structure across views. In particular, we propose a general optimization problem, where graph signals are assumed to be smooth over the unknown multiview graph. The proposed framework employs a consensus-based regularization, where the views are encouraged to be similar to each other through a learned \textit{consensus graph}, which represents the shared structure across views. By learning the consensus graph, our method can reveal the shared structure across views, which can also be used to identify the unique features of each view. The proposed optimization problem is written in a general form that allows different regularization functions to be employed without changing the optimization procedure. One can choose these functions based on how views are related to each other and what the shared structure across views is. In order to illustrate the flexibility of our method, we propose two regularization functions that are inspired by fused and group graphical lasso \cite{danaher2014joint}. 

To summarize, the contributions of our work are:
\begin{itemize}
    \item Extending smoothness based graph learning to multiview graph setup through consensus-based regularization, where both the individual view graphs and the consensus graph, representing the shared structure across views, are learned. 
    \item A general optimization framework that allows using different regularization functions which can be selected based on how views are related to each other and what the shared structure across views is. 
    \item Two multiview graph learning models that assume the topological similarity of views. Proposed models differ from each other based on how much they allow the views to vary from the consensus graph.
\end{itemize}
The remaining of the paper is organized as follows. In Section \ref{sec:related_work}, a review of related work is provided. Section \ref{sec:background} provides a background on graphs and graph learning. Section \ref{sec:mvgl} describes the proposed multiview graph learning. Section \ref{sec:results} includes results on different simulated data as well as on functional brain connectivity network inference from EEG recordings of multiple subjects.
\section{Related Work}
\label{sec:related_work}
Most previous methods for joint inference of views of a multiview graph are based on statistical models. These methods extend graphical lasso \cite{friedman2008sparse} to joint learning setup, where they learn the precision matrices of multiple related Gaussian graphical models and use various penalties in the likelihood framework to exploit the common characteristics shared by different views \cite{guo2011joint, danaher2014joint, lee2015joint, mohan2014node, ma2016joint, huang2015joint}. Most notable work using this approach is the joint graphical lasso \cite{danaher2014joint}, where fused or group lasso penalties are used to encourage topological similarity across views. However, these methods are limited by the assumption that the observed graph signals are Gaussian which is usually not true for real-world applications. They also focus on learning the precision matrices without imposing any graph structure constraints on the learned views. Therefore, they do not infer the graph structure but rather learn conditional dependencies, which may not be suitable for interpreting the structure of data in some contexts. Recently, these joint learning approaches have been extended to jointly learn multiple graph Laplacian matrices instead of precision matrices \cite{yuan2023joint}. However, this approach is still limited to Gaussian data. Finally, none of the aforementioned approaches aims to learn a consensus graph, thus they cannot reveal the shared structure across views.

Recently, GSP community has addressed the problem of learning multiview graphs from heterogeneous data. This work can be split into two categories depending on whether one knows the association of the observed signals to the views \textit{a priori}. In the first setup, multiple datasets are given and each dataset is defined on a view \cite{navarro2022joint, navarro2022jointb}. On the other hand, the second setup deals with the mixture of graph signals, where one is given a single dataset and the association of graph signals to the views is not known \cite{maretic2020graph, araghi2019k, karaaslanli2022simultaneous}. The focus of the present paper is the first category. This problem setting has been most widely studied for inferring the topology of time-varying networks \cite{kalofolias2017learning, yamada2019time, baingana2016tracking, sardellitti2019enabling}, where the aim is to learn graphs at multiple time points and to track changes in the graph structure across time. This problem can be posed as multiview graph learning with a regularization term that promotes pre-specified changes between consecutive graphs. More recently, the problem of multiview graph learning has been formulated with the assumption of graph stationarity \cite{navarro2022joint}. In this formulation, the signals are assumed to be stationary and pairwise similarity between all graphs is used to regularize the optimization. While this formulation has the same goal as the present paper, it is based on stationary graph signals; while we focus on smoothness. Moreover, it does not learn a consensus graph. \cite{zhang2024graph} propose a multiview graph learning method based on smoothness assumption. However, it focuses on decentralized optimization by relying on pairwise regularization, which does not allow for the inference of the shared structure across views.
\section{Background}
\label{sec:background}
\subsection{Notations}
In this paper, lowercase and uppercase letters, \textit{e.g.}, $n$ or $N$, are used to represent scalars. Vectors and matrices are shown as lowercase and uppercase bold letters, \textit{e.g.}, $\vx$ and $\mX$, respectively. For a vector $\vx$, $x_{i}$ is its $i$th entry. For a matrix $\mX$, we use $X_{ij}$, $\mX_{i\cdot}$ and $\mX_{\cdot i}$ to show its $ij$th entry, $i$th row and $i$th column, respectively. Vectors with all entries equal to 1 or 0, and identity matrix are shown as $\ones$, $\zeros$, and $\mI$, respectively. For a vector $\vx \in \setR^{N}$, $\diag(\vx)$ is a diagonal matrix $\mX \in \setR^{N \times N}$ with $X_{ii} = x_i$. For a matrix $\mX \in \setR^{N \times N}$, $\diag(\mX)$ is a vector $\vx$ with $x_{i} = X_{ii}$. The operator $\upper(\cdot)$ takes a symmetric matrix $\mX \in \setR^{n\times n}$ as input and returns its upper triangular part as a vector $\vx \in \setR^{n(n-1)/2}$ constructed in row-major order. We define the matrix $\mS \in \setR^{n \times n(n-1)/2}$ such that $\mS \upper(\mX) = \mX \ones$. 

\subsection{Graphs and Graph Signals}
An undirected weighted graph is represented as $G=(V, E, \mW)$ where $V$ is the node set with cardinality $n$, $E$ is the edge set. $\mW \in \setR^{n\times n}$ is the adjacency matrix of $G$, where $W_{ij} = W_{ji}$ is the weight of the edge between nodes $i$ and $j$ and $W_{ij} = W_{ji} = 0$ if there is no edge between nodes $i$ and $j$. $\vd = \mW \ones$ is the degree vector and $\mD = \diag(\vd)$ is the diagonal degree matrix. The Laplacian matrix of $G$ is defined as $\mL = \mD - \mW$. Its eigendecomposition is $\mL = \mV^\top \mLambda \mV$, where columns of $\mV$ are eigenvectors and $\mLambda$ is the diagonal matrix of eigenvalues with $0 = \Lambda_{11} \leq \Lambda_{22} \leq \dots \leq \Lambda_{nn}$.

A graph signal defined on $G$ is a function $x : V \to \setR$ and can be represented as a vector $\vx \in \setR^{n}$ where $x_{i}$ is the signal value on node $i$. Eigenvectors and eigenvalues of the Laplacian of $G$ can be used to define the graph Fourier transform, where small eigenvalues correspond to low frequencies. Thus, the graph Fourier transform of $\vx$ is $\widehat{\vx} = \mV^\top \vx$ where $x_{i}$ is the Fourier coefficient at the $i$th frequency component $\Lambda_{ii}$. $\vx$ is a smooth graph signal, if most of the energy of $\widehat{\vx}$ lies in low frequency components. We can calculate the total variation of $\vx$ as:
\begin{align}
    \label{eq:smoothness}
    \trace(\widehat{\vx}^\top \mLambda \widehat{\vx}) = \trace(\vx^\top \mV \mLambda \mV^T \vx) = \trace(\vx^\top \mL \vx).
\end{align}
\subsection{Single Graph Learning}
An unknown graph $G$ can be learned from a set of graph signals defined on it based on some assumptions about the relation between the observed graph signals and the underlying graph structure. Dong et. al. \cite{dong2016learning} proposed to learn $G$ by assuming the graph signals are smooth with respect to $G$. Given $\mX \in \setR^{n \times p}$ as the data matrix with the columns corresponding to the observed graph signals, $G$ can be learned by minimizing smoothness, defined in \eqref{eq:smoothness}, with respect to the Laplacian matrix of $G$:
\begin{align}
\label{eq:single_graph_learning}
\begin{aligned}
    \minimize_{\mL} &\ \trace({\mX}^\top \mL \mX) + \alpha \norm{\mL}_F^2 \\
    \textrm{s.t.}\quad\ &\  \mL \in \setL \textrm{ and } \trace(\mL) = 2n,
\end{aligned}
\end{align}
where the first term quantifies the total variation of graph signals and the second term controls the density of the learned graph such that larger values of hyperparameter $\alpha$ result in a denser graph. $\mL$ is constrained to be in $\setL=\{\mL : L_{ij} = L_{ji} \leq 0\ \forall i\neq j,\ \mL \ones = \zeros\}$, which is the set of valid Laplacians. The second constraint is added to prevent the trivial solution $\mL = \zeros$.

\section{Multiview Graph Learning}
\label{sec:mvgl}
\subsection{Problem Formulation}
\label{ssec:problem_formulation}
In multiview graph learning, the aim is to jointly learn $N$ closely related graphs, $\calG=\{G^1, \dots, G^N\}$. The graphs are defined as $G^i = (V, E^i, \mW^i)$ with $|V|=n$ and are assumed to share a common structure represented by a \textit{consensus graph} $G = (V, E, \mW)$. Our aim is to learn both graphs in $\calG$ and the consensus graph $G$ from the collection of data matrices, $\calX = \{\mX^1, \dots, \mX^N\}$, where the columns of $\mX^i \in \setR^{n \times p_i}$ are assumed to be smooth graph signals defined on $G^i$. In particular, let $\mL^i$ and $\mL$ be the graph Laplacians of $G^i$ and $G$, respectively. Graph Laplacians can then be jointly learned by the following optimization problem:
\begin{align}
\label{eq:multiview_graph_learning}
    \minimize_{\{\mL^i\}_{i=1}^N, \mL} 
        & \sum_{i=1}^N 
            \big\{\trace({\mX^i}^\top \mL^i \mX^i) + 
            \alpha \norm{\mL^i}_F^2\big\} \notag \\ 
        & + \beta c(\{\mL^i - \mL\}_{i=1}^N) + 
            \gamma r(\mL) \\
    \textrm{s.t.}\quad\ 
        &\ \mL^i \in \setL, \trace(\mL^i) = 2n, \forall i, 
        \textrm{ and } \mL \in \setL, \notag
\end{align}
where the first sum and constraints are analogous to \eqref{eq:single_graph_learning}. Since $G$ is the structure shared by $G^i$'s, learned $\mL^i$'s are encouraged to be similar to $\mL$ through $c(\{\mL^i - \mL\}_{i=1}^N)$, which penalizes the difference between $\mL^i$'s and $\mL$. $r(\mL)$ is added to regularize the learned $\mL$ to have the desired properties. Finally, $\alpha$, $\beta$ and $\gamma$ are hyperparameters that scale different terms of the objective function. Before giving details about the forms of $c(\cdot)$ and $r(\cdot)$, we propose an optimization algorithm that can be used to solve  \eqref{eq:multiview_graph_learning} irrespective of the forms of $c(\cdot)$ and $r(\cdot)$.
\subsection{Optimization} 
\label{ssec:optimization}
In order to solve the optimization problem in \eqref{eq:multiview_graph_learning}, we first vectorize it such that the upper triangular parts of the graph Laplacians are learned. Let $\vk^i = \upper(\mX^i{\mX^i}^\top) \in \setR^m$, $\vd^i=\diag(\mX^i{\mX^i}^\top) \in \setR^n$, $\vell^i = \upper(\mL^i) \in \setR^m$ and $\vell = \upper(\mL) \in \setR^m$, where $m = n(n-1)/2$. Also, define functions $c_v(\{\vell^i - \vell\}_{i=1}^N) = c(\{\mL^i - \mL\}_{i=1}^N)$ and $r_v(\vell) = r(\mL)$. The problem in \eqref{eq:multiview_graph_learning} can then be vectorized as follows (see Appendix \ref{appdx:vectorization} for details):
\begin{align}
\label{eq:multiview_graph_learning_vec}
    \minimize_{\{\vell^i\}_{i=1}^N, \vell} 
        & \sum_{i=1}^N \big\{
            (2\vk^i - \mS^\top\vd^i)^\top \vell^i +
            \alpha {\vell^i}^\top(\mS^\top \mS + 2\mI)\vell^i 
        \big\} \notag \\
        & + \beta c_v(\{\vell^i - \vell\}_{i=1}^N)
        + \gamma r_v(\vell) \\
    \textrm{s.t.}\quad\ & \ \vell^i \leq 0, \ones^\top \vell^i = -n, \forall i, \textrm{ and } \vell \leq 0, \notag
\end{align}
which can be solved using Alternating Direction Method of Multipliers (ADMM) \cite{boyd2011distributed}. Let $f(\vell^i) = (2\vk^i - \mS^\top\vd^i)^\top \vell^i + \alpha {\vell^i}^\top(\mS^\top \mS + 2\mI)\vell^i$, and let $\imath_1(\cdot)$ and $\imath_2(\cdot)$ be the indicator functions for the sets $\{\vell \in \setR^m | \ones^\top \vell = -n\}$ and $\{\vell \in \setR^m | \vell \leq 0\}$, respectively. We introduce the slack variables $\vz = \vell,\ \vz^i = \vell^i,\ \vv^i=\vz^i - \vz\ \forall i$ and rewrite the problem in its standard ADMM form as follows:
\begin{align}
\begin{aligned}
\label{eq:multiview_graph_learning_admm}
    \minimize_{\{\vell^i, \vz^i, \vv^i\}_{i=1}^N, \vell, \vz} 
        & \sum_{i=1}^N \big\{
            f(\vell^i) + \imath_1(\vell^i) + \imath_2(\vz^i) 
        \big\} \\
        & + \beta c_v(\{\vv^i\}_{i=1}^N) + \gamma r_v(\vell) + \imath_2(\vz) \\
    \textrm{s.t.} \quad \quad \ 
        & \ \vell^i = \vz^i, \vv^i = \vz^i - \vz\ \forall i,  \textrm{ and } \vell = \vz,
\end{aligned}
\end{align}
where the first and last constraints in \eqref{eq:multiview_graph_learning_vec} are imposed onto $\vz^i$'s and $\vz$, respectively and all constraints are included in the objective function through the indicator functions. The augmented Lagrangian can then be written as:
\begin{align}
\label{eq:augmented_lagrangian}
    \calL(\{&\vell^i, \vz^i, \vv^i, \vy^i, \vw^i\}_{i=1}^N, \vell, \vz, \vy) \notag \\
        = & \begin{aligned}[t] \sum_{i=1}^N \Big\{ & 
                f(\vell^i) + \imath_1(\vell^i) + \imath_2(\vz^i) 
                + {\vy^i}^\top(\vz^i - \vell^i) \\
                + & \frac{\rho}{2} \norm{\vz^i - \vell^i}_2^2 
                + {\vw^i}^\top (\vv^i - \vz^i + \vz) \\
                + & \frac{\rho}{2} \norm{\vv^i - \vz^i + \vz}_2^2 
        \Big\} + \beta c_v(\{\vv^i\}_{i=1}^N) \end{aligned} \notag \\
    & + \gamma r_v(\vell) + \imath_2(\vz) + \vy^\top(\vz-\vell) + \frac{\rho}{2} \norm{\vz-\vell}_2^2,
\end{align}
where $\vy^i$'s, $\vw^i$'s and $\vy$ are the Lagrangian multipliers of the equality constraints in \eqref{eq:multiview_graph_learning_admm} and $\rho$ is the penalty parameter. Standard ADMM steps at the $k$th iteration are:
\begin{align}
    \{\currval{\vell}^i, \currval{\vv}^i\}_{i=1}^N, \currval{\vell} & = \argmin_{\{\vell^i, \vv^i\}_{i=1}^N, \vell} \calL(\{\vell^i, \prevval{\vz}^i, \vv^i, \prevval{\vy}^i, \prevval{\vw}^i\}_{i=1}^N, \vell, \prevval{\vz}, \prevval{\vy}) \label{eq:admm_first} \\
    \{\currval{\vz}^i\}_{i=1}^N, \currval \vz & = \argmin_{\{\vz^i\}_{i=1}^N, \vz} \calL(\{\currval{\vell}^i, \vz^i, \currval{\vv}^i, \prevval{\vy}^i, \prevval{\vw}^i\}_{i=1}^N, \currval{\vell}, \vz, \prevval{\vy}) \raisetag{2em} \label{eq:admm_second}  \\
    \currval{\vy}^i & = \prevval{\vy}^i + \rho(\currval{\vz}^i - \currval{\vell}^i),\ \forall i \\
    \currval{\vw}^i & = \prevval{\vw}^i + \rho(\currval{\vv}^i - \currval{\vz}^i + \currval{\vz}),\ \forall i \\
    \currval{\vy} & = \prevval{\vy} + \rho(\currval{\vz} - \currval{\vell}),
\end{align}
where variables with $\currval{\phantom{x}}$ and $\prevval{\phantom{x}}$ represent the values of those variables at $k$th and $(k-1)$th iteration, respectively. The problem in \eqref{eq:admm_first} is separable across its variables, thus it can be solved with respect to $\vell^i$'s, $\vv^i$'s and $\vell$ separately. Its optimization with respect to each $\vell^i$ leads to a quadratic problem with equality constraint, which has a closed form solution derived from the corresponding Karush–Kuhn–Tucker (KKT) conditions. Optimizing \eqref{eq:admm_first} with respect to  $\vv^i$'s results in the proximal operator of $c_v$. Similarly, solution of \eqref{eq:admm_first} with respect to $\vell$ is the proximal operator of $r_v$. The variables of \eqref{eq:admm_second} are coupled, therefore it cannot be solved separately for each $\vz^i$ and $\vz$. We solve it using block coordinate descent (BCD) as it's a smooth convex problem with separable inequality constraints \cite{shi2016primer}. The details of the solutions for \eqref{eq:admm_first} and \eqref{eq:admm_second} are given in Appendix \ref{appdx:admm_steps}.
\subsection{Selection of $c(\cdot)$ and $r(\cdot)$}
\label{ssec:model_selection}
Different forms for $c(\cdot)$ and $r(\cdot)$ can be selected depending on what the shared structure across $G^i$'s is and what properties are desired for $G$.  In this section, we propose two models in order to show that the proposed optimization problem and its solution are generalizable to  different scenarios. For ease of interpretation, the models are defined using the vectorized form of the regularization terms, i.e. $c_v(\cdot)$ and $r_v(\cdot)$. 

The first model, which is referred as \mvglone, is defined as follows:
\begin{align}
        \sum_{i=1}^N c_v(\vell^i - \vell) & = \sum_{i=1}^N \norm{\vell^i - \vell}_1 = \norm{\Delta}_{1, 1}, \label{eq:case1_c}\\
        r_v(\vell) & = 0 \label{eq:case1_r},
\end{align}
where $\Delta \in \setR^{m \times N}$ with $\Delta_{\cdot i} = \vell^i - \vell$, $m=n(n-1)/2$, and $\norm{\cdot}_{p,q}$ is the element-wise $\ell_{p,q}$ matrix norm. This model can be considered as analogous to fused graphical lasso \cite{danaher2014joint} that extends fused lasso regularization \cite{tibshirani2005sparsity} to graphical lasso problem. As in fused graphical lasso, \mvglone{} encourages the topology of the different views to be similar to each other. However, unlike fused graphical lasso which focuses on the pairwise differences, \mvglone{} ensures this goal through the consensus graph, which can be considered as the common structure shared by the views. Moreover, if an edge differs across views, it guarantees that only a few views differ as it applies $\ell_1$ norm regularization across the views and the edges simultaneously through $\ell_{1,1}$ norm. 

This property can be restrictive in some situations; thus, we define a second model referred to as \mvgltwo:
\begin{align}
    \sum_{i=1}^N c_v(\vell^i - \vell) & = \sum_{a=1}^m \sqrt{\sum_{i=1}^N (\ell_a^i - \ell_a)^2} = \norm{\Delta}_{2, 1}, \label{eq:case2_c} \\ 
    r_v(\vell) & = \norm{\vell}_1 \label{eq:case2_r},
\end{align}
which can be seen as the extension of group graphical lasso \cite{danaher2014joint} to the case of consensus based multiview graph learning. As in the first model, \mvgltwo{} ensures the topologies to be similar across views. However, if an edge differs across views, all views are allowed to differ as it utilizes $\ell_{2,1}$ matrix-norm \cite{yang2011l21}. Moreover, $\vell$ is regularized with $\ell_1$ norm to impose sparsity. Compared to the first model, this is necessary; since without this regularization $\vell$ is equivalent to the element-wise mean of the views, which results in a dense consensus graph. Therefore, we employ $\ell_1$ norm regularization to impose sparsity on the learned $\vell$. On the other hand, in \mvglone, $\vell$ is the element-wise median of the views  which ensures that the learned $\vell$ has a sparsity level similar to the individual views. Thus, a sparsity imposing regularization is not necessary.
\subsection{Convergence and Computational Complexity}
Our ADMM based optimization procedure is derived by rewriting the vectorized problem \eqref{eq:multiview_graph_learning_vec} as in \eqref{eq:multiview_graph_learning_admm} using slack variables. This new form of the problem follows the standard two-block ADMM form, therefore it converges (see \cite{boyd2011distributed, eckstein2015understanding}). The computational complexity of the procedure can be calculated by considering the ADMM steps. Namely, the solution of \eqref{eq:admm_first} with respect to each $\vell^i$ is $\calO(n^2)$ where $n$ is the number of nodes. Complexity of solving \eqref{eq:admm_first} with respect to $\vv^i$'s and $\vell$ depends on the proximal operator of $c_v$ and $r_v$. For the two models proposed in Section \ref{ssec:model_selection}, the time complexity is $\calO(n^2N)$ where $N$ is the number of views. Each subproblem of the proposed BCD procedure for solving \eqref{eq:admm_second} are projections onto the non-positive orthant with time complexity of $\calO(n^2)$. If the number of BCD iterations is $I_1$, solving \eqref{eq:admm_second} costs $\calO(n^2NI_1)$. In conclusion, assuming $I_2$ is the number of ADMM iterations, the time complexity of our optimization is $\calO(n^2NI_1I_2)$ when $c_v$ and $r_v$ are selected as described in section \ref{ssec:model_selection}.
\begin{figure*}[t]
    \centering%
    \includegraphics[width=\textwidth]{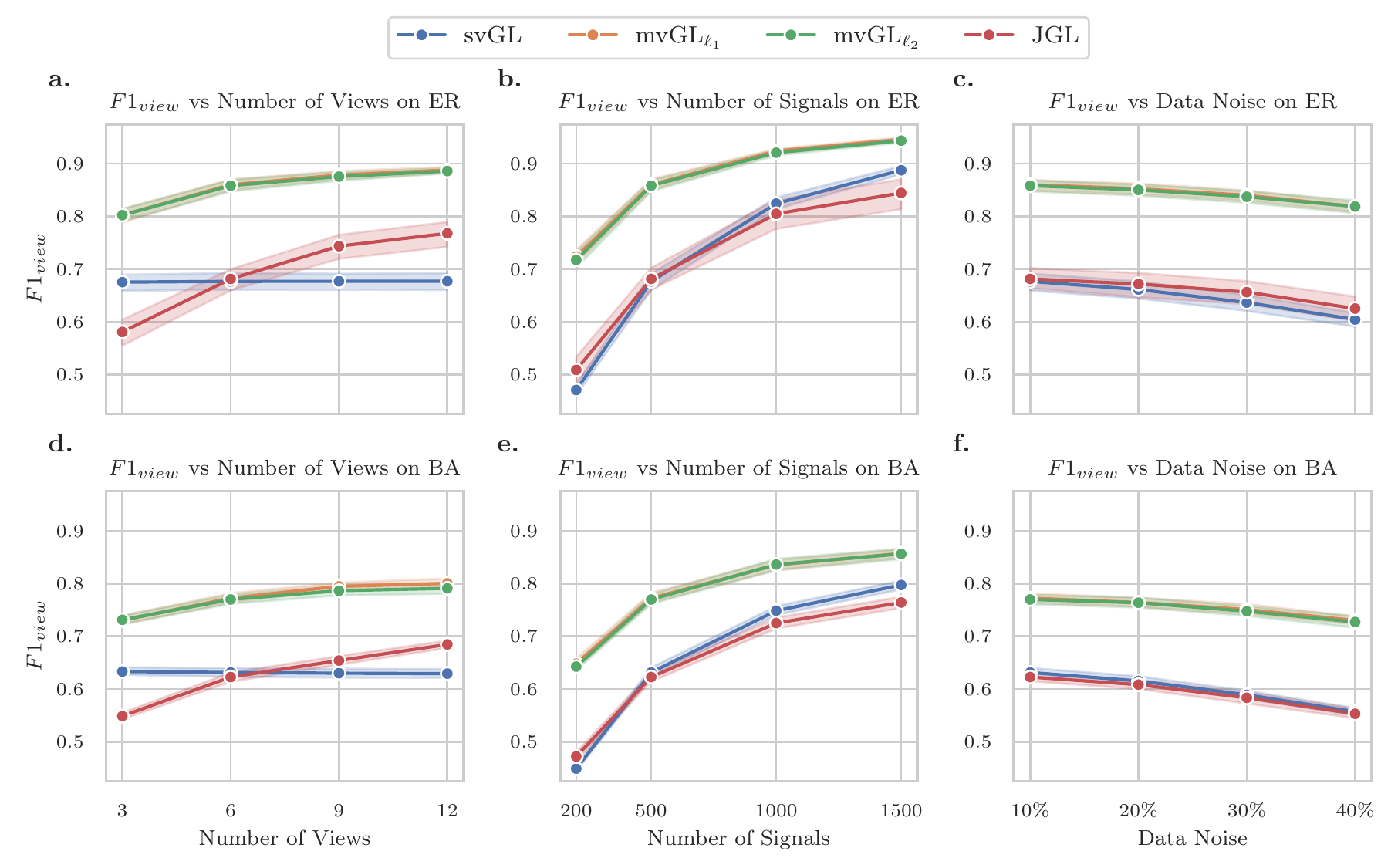}%
    \caption{Performance of methods when learning view graphs. a., b. and c. show the results when the consensus graph is generated using \ermodel{}. d., e. and f. show the performance when the consensus graph is drawn from \bamodel{}. Changes in performance with respect to number of views (a. and d.), number of signals (b. and e.) and the amount on noise (c. and f.) are plotted. Shaded area is the $95\%$ confidence interval calculated from $10$ realizations of simulated data.}%
    \label{fig:simulations_view_performance}%
\end{figure*}
\section{Results}
\label{sec:results}
In this section, the proposed multiview graph learning approach is tested on both simulated data and applied to a real world application of inferring functional connectivity networks of multiple subjects from EEG recordings. 
\subsection{Simulated Datasets}
The proposed algorithms are applied to simulated data where $\calX$ is generated from multiple related graphs whose structures are known. The learned graphs are then compared to the ground truth graphs to evaluate the performance. The proposed methods are compared against single view graph learning (svGL) and fused joint graphical lasso (JGL). For svGL, we optimize \eqref{eq:single_graph_learning} using ADMM for each dataset in $\calX$ such that each graph is learned separately. Therefore, svGL provides the baseline performance, as it does not share information across the views during learning. JGL learns multiple related precision matrices of Gaussian graphical models. In particular, we compare with fused JGL where each pair of precision matrices are regularized to be similar to each other through $\ell_1$ norm. 

\paragraph*{Data Generation} Simulated data are generated from known graphs that are constructed as follows. We first generate a graph $G$ with $n$ nodes from two random graph models: \ermodel{} model with edge probability $0.1$ and \bamodel{} model with the growth parameter $m=5$. Next, $N$ view graphs are generated, where each $G^i$ is generated independently from $G$ by randomly shuffling $10\%$ of $G$'s edges. In this procedure, $G$ is considered as the consensus graph for the view graphs $\calG = \{G^1, \dots, G^N\}$. Given the view graphs, each $\mX^i \in \setR^{n \times p}$ in $\calX$ is generated from $G^i$ using the smooth graph filter $h(\mL^i)$. Namely, each column of $\mX^i$ is generated as $\mX_{\cdot j}^i = h(\mL^i) \vx_0$; where $h(\mL^i) = {\mL^i}^\dagger$, $\vx_0 \sim \calN(\zeros, \mI)$ and $^\dagger$ is the pseudo-inverse. We finally add $\eta\%$ noise (in $\vell_2$ norm sense) to each generated $\mX^i$.  For each simulation, average performance over 10 realizations is reported.
\begin{figure*}[b]
    \centering
    \includegraphics[width=\textwidth]{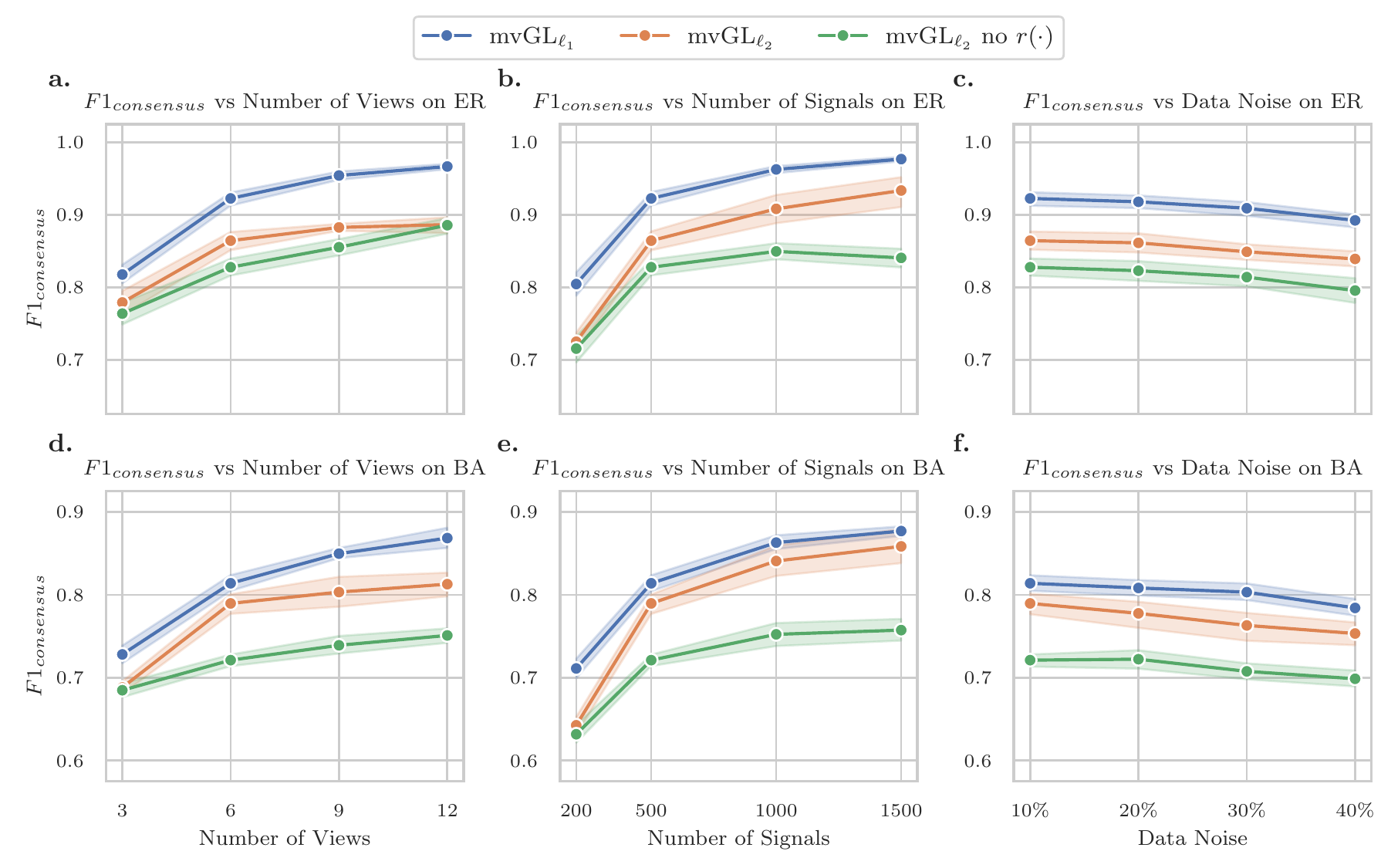}
    \caption{Performance of methods when learning the consensus graphs. a., b. and c. show the results when consensus graph is generated with \ermodel{}; and d., e. and f. show the performance for \bamodel{} model. Changes in performance with respect to number of views (a. and d.), number of signals (b. and e.) and the amount on noise (c. and f.) are plotted. Shaded area is the $95\%$ confidence interval calculated from $10$ realizations of simulated data.}
    \label{fig:simulations_consensus_performance}
\end{figure*}
\paragraph*{Performance Metric} We quantify the performance of the different methods using F1 score, which measures how well the learned edges match with the true edges. In particular, let $\widehat{\calG} = \{\widehat{G}^1, \dots, \widehat{G}^N\}$ and $\calG = \{G^1, \dots, G^N\}$ be the learned and the ground truth graph structures, respectively. We measure the performance of different methods using F1 score averaged over the views, i.e. $F1_{\text{view}} = \frac{1}{N} \sum_{i=1}^N F1(G^i, \widehat{G}^i)$. The performance of the proposed methods in inferring the consensus graph is also calculated as $F1_{\text{consensus}} = F1(G, \widehat{G})$ where $G$ and $\widehat{G}$ are the ground truth and learned consensus graphs, respectively.
\paragraph*{Hyperparameter Selection} Proposed methods, svGL and JGL, require the selection of hyperparameters that weigh the different parts of their loss functions. For \mvglone{} and \mvgltwo{}, $\alpha$ parameter determines the sparsity of the learned view graphs. Similarly, svGL and fused JGL have analogous parameters that determine the sparsity of the  graphs. For all the methods, we perform a grid search on these parameters and use the value that provides the best $F1_{\text{view}}$ value. For \mvgltwo{}, $\gamma$ parameter determines the sparsity of the learned consensus graph. $\gamma$ is also optimized to find the value that gives the best $F1_{\text{view}}$. The joint learning methods, i.e. the proposed methods and JGL, also require the selection of a parameter which tunes how similar the learned view graphs are to each other (parameter $\beta$ in \eqref{eq:multiview_graph_learning}). In all experiments, we search for the value where the correlation between the adjacency matrices of the learned view graphs is approximately $0.8$. We selected this value by calculating the correlation between view graphs when they are learned with svGL. As svGL does not impose any similarity across graphs, our goal is to ensure that the correlation between the view graphs is higher when they are learned jointly. We find $0.8$ as the value that satisfies this condition in all considered experiments. 
\paragraph*{Performance of View Graph Learning} Methods are first compared based on their performance on learning the view graphs. We evaluate the performance of different methods with respect to varying number of views $N$, number of signals $p$ and the amount of noise $\eta$.

Figures \ref{fig:simulations_view_performance}a. and \ref{fig:simulations_view_performance}d. show the performance when $N$ is increased from 3 to 12, while the number of nodes $n$ is 100, $p=500$ and $\eta = 0.1$ for \ermodel{} and \bamodel{} models, respectively. In both plots, the proposed models have very similar performance. This behaviour could be due to the fact that both models assume the views have similar topologies and the view graphs in our simulated data are inline with this assumption. Both models have higher performance than svGL, indicating that the proposed optimization problem in \eqref{eq:multiview_graph_learning} can effectively share information across views irrespective of which $c(\cdot)$ is used. JGL shows poor performance compared to the proposed models. Its performance is lower than or similar to svGL for $N=3$ and $N=6$ and it performs better than svGL for higher values of $N$. The poor performance of JGL could be due to the fact that it is designed to learn precision matrices rather than Laplacian matrices. Finally, the figures reveal that the performance of joint learning methods improves with increasing number of views; which is expected as more information is available to learn from. On the other hand, svGL is not affected by the increase in the number of views, since it does not share information across views when learning graphs. 

In Figures \ref{fig:simulations_view_performance}b. and \ref{fig:simulations_view_performance}e., the performance of methods with respect to $p$ is shown with $n=100$, $N=6$, and $\eta = 0.1$ for \ermodel{} and \bamodel{} models, respectively.  As above, the proposed models have similar performance and their $F1_{view}$ scores are higher than those of svGL and JGL. As expected, the performance of all methods improves with increasing number of signals. Figures \ref{fig:simulations_view_performance}c. and \ref{fig:simulations_view_performance}f. show the effect of noise on methods for $n = 100$, $p = 500$ and $N = 6$. The comparison between the methods is the same as the previous two cases. We observe a drop in the performance of all methods with increasing data noise. 
\paragraph*{Performance of Consensus Graph Learning} In this experiment, \mvglone{} and \mvgltwo{} are compared based on their performance in learning the consensus graph $G$. We also compare both models to a modified version of \mvgltwo{}, where the $r(\cdot)$ term is removed, which allows us to show the importance of regularization for consensus graph learning. Finally, svGL and JGL are not considered for comparison as they do not learn a consensus graph. As before, we vary the number of views, number of signals and the amount of noise added to signals during simulated data generation and observe how these changes affect the models. In all cases, the remaining parameters of data generation process are set as before. 

\begin{figure}[t]
    \centering
    \includegraphics[width=\columnwidth]{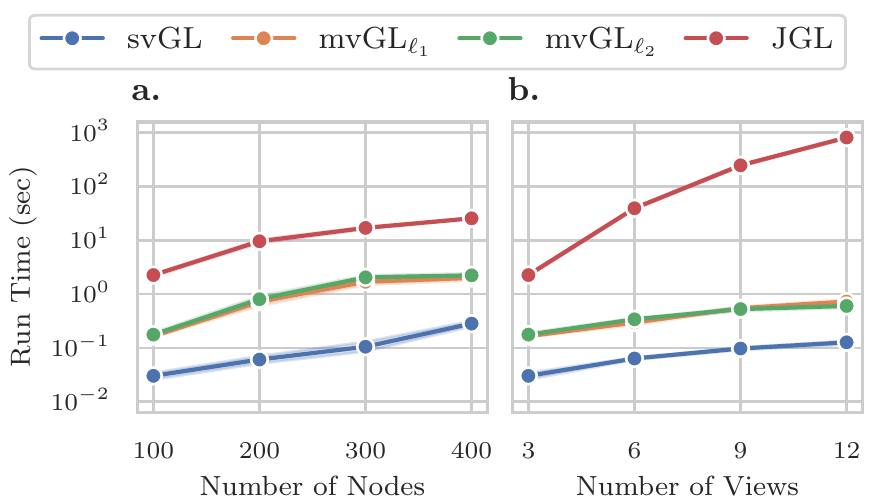}
    \caption{Run times of methods with increasing number of nodes (a.) and number of views (b.). For svGL, we report total run time to learn all views. }
    \label{fig:time_complexity}
\end{figure}

Figure \ref{fig:simulations_consensus_performance} shows the results, where we observe that \mvglone{} is better than \mvgltwo{} in identifying the consensus graph. The reason for this is that the relation between ground truth view graphs and the consensus graph in the simulated data is more inline with \mvglone{} assumptions. Namely, each view is \textit{independently} generated from the consensus graph by shuffling a fraction of randomly selected edges in the consensus graph. Thus, if an edge differs across views, only a few number of views are different than the consensus graph. Note that this does not mean \mvgltwo{} is a worse model compared to \mvglone{}. It is only the case that \mvglone{} is the better model for this dataset and \mvgltwo{} can have better performance in other situations. Figure \ref{fig:simulations_consensus_performance} also shows that including the $r(\cdot)$ term in \mvgltwo{} improves its performance; since otherwise, the learned consensus graph is denser and includes more erroneous edges. This result indicates the importance of regularizing the consensus graph and the flexibility of the proposed optimization model in \eqref{eq:multiview_graph_learning}. Finally, similar to view performance, increasing the number of views and signals leads to better inference; while increasing the data noise reduces the performance.
\paragraph*{Time complexity comparison} Lastly, we compare the methods based on their time complexity, where all methods are run on the same compute clusters with the same resource allocation. Run times are measured on simulated data, where the consensus graph is generated using \ermodel{} model with edge probability set to 0.1. Figure \ref{fig:time_complexity}a. shows the run times in seconds as a function of number of nodes, where the remaining parameters of simulated data generation are set as $N=3$, $\eta = 0.1$ and $p=500$. All methods are observed to have longer run time with increasing number of nodes due to increased number of unknowns. svGL is the fastest method, which is expected as it infers each view separately without any regularization which makes it an easier optimization problem. svGL is followed by the proposed models, which have the same time complexity. JGL is found to be the slowest method since its computational complexity is cubic in $n$, while the other methods are quadratic in $n$. In Figure \ref{fig:time_complexity}b., run times as a function of the number of views are shown, where the remaining parameters of the simulations are $n=100$, $\eta = 0.1$ and $p = 500$. As in Figure \ref{fig:time_complexity}a., the fastest method is svGL and the slowest one is JGL. As expected, run times of all methods increases with number of views. 

\subsection{EEG Dataset}
In this section, the proposed method is applied to EEG recordings collected from multiple subjects, where the goal is to learn the functional connectivity  networks (FCNs) for each subject. Since the subjects perform the same cognitive task, it is assumed that there is a common network structure across subjects. By applying the proposed methods, the learned FCNs are ensured to be similar to each other and the learned consensus graph represents the common structure across subjects. The graphs learned by the proposed framework are validated by (i) comparing them to those found by svGL and (ii) comparing their topology to existing literature. 
\paragraph{Data Description} The EEG data is recorded from 20 subjects performing a cognitive control-related error processing task \cite{moran2012sex}. The data collection was conducted by following the experimental protocol approved by the Institutional Review Board (IRB) of the Michigan State University (IRB: LEGACY13-144). The EEG signals are recorded with a BioSemi ActiveTwo system using a cap with 64 Ag–AgCl electrodes placed at standard locations of the International 10–20 system. The sampling rate was 512 Hz. The collected data is processed using standard artifact rejection algorithms \cite{delorme2004eeglab}, followed by the volume conduction minimization from the Current Source Density (CSD) Toolbox \cite{tenke2012generator}.
\begin{figure*}[t]
    \centering
    \includegraphics[width=0.8\textwidth]{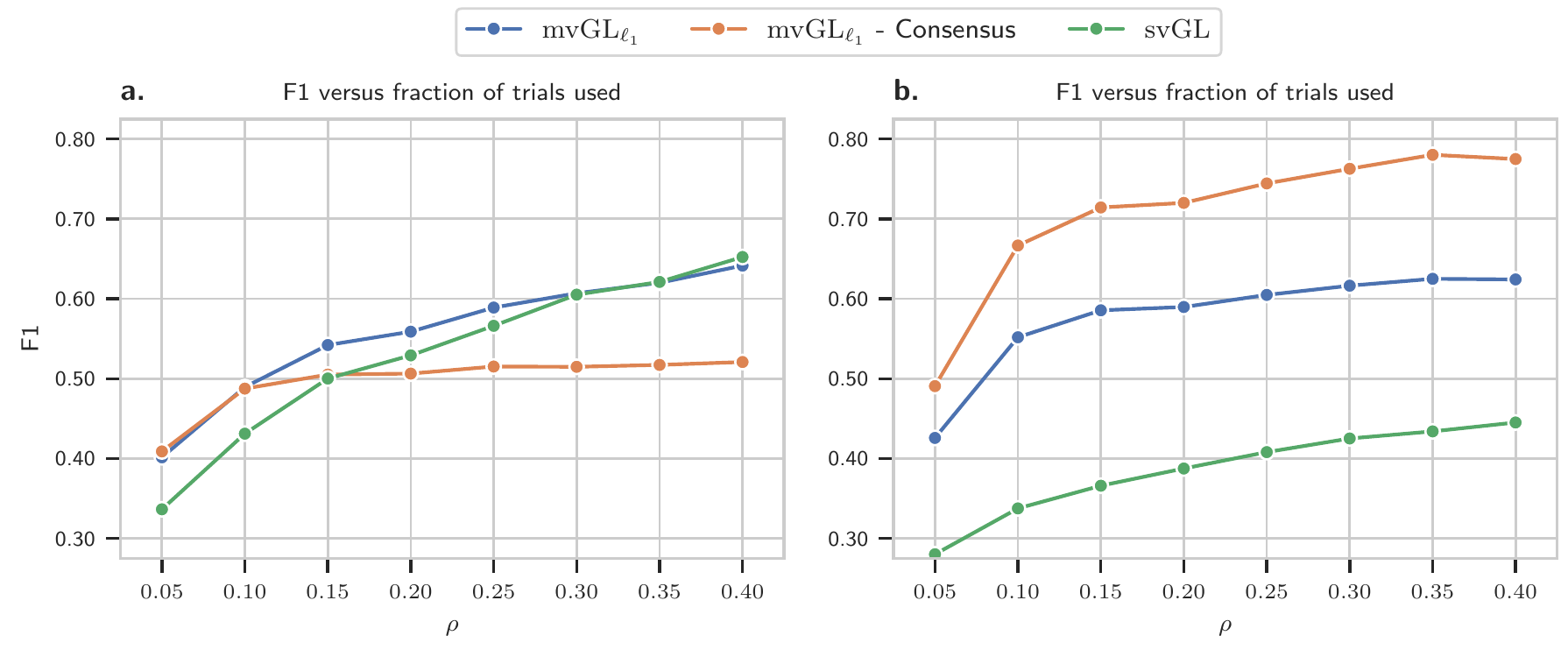}%
    \caption{Performance of different methods when "ground-truth" graphs are constructed with \svgl{} using all EEG trials. a) A "ground-truth" graph is constructed for each subject using all available trials of the subject. b) A single "ground-truth" graph is constructed across all trials and subjects.}
    \label{fig:eeg_gt_based_comparison}
\end{figure*}
During recordings, subjects perform a letter version of the speeded reaction Flanker task, where a string of five letters, either congruent (e.g., SSSSS) or incongruent (e.g., SSTSS), is shown at each trial. Subjects use a standard mouse to react to the center letter and the goal is to capture the Error-Related Negativity (ERN) after an error response. Each trial begins with a flanking stimulus (e.g., SS SS) of 35ms followed by the target stimuli (e.g., SSSSS/SSTSS) displayed for about 100 ms. There is a 1200 to 1700 ms break between trials. Each subject performs 480 trials, where number of trials with error response ranges from 34 to 139 across subjects. In this paper, data from error trials are employed for graph learning and only 34 error trials corresponding to each subject are taken into account to ensure that the subjects have the same amount of data. 
\paragraph{Preprocessing} Before performing graph learning, we preprocess the EEG data as follows. Let $\overline{\mX}^{sk} \in \setR^{64\times 512}$ be the EEG recordings of the $s$th subject's $k$th trial, where $64$ is the number of electrodes and $512$ is the number of time points. Thus, each row of $\overline{\mX}^{sk}$ corresponds to EEG signals from an electrode. Based on our previous work \cite{aviyente2011phase}, which indicates increased error related activity for the $\theta$ frequency band (4-7 Hz) and time window (approximately 0-100 ms), we first bandpass filter each row of $\overline{\mX}^{sk}$ within the $\theta$ band and then consider samples from 0-100 ms. We perform graph learning using  the data matrix, $\mX^{sk} \in \setR^{64\times 50}$, obtained after filtering and windowing operations.

\paragraph{"Ground-Truth" based comparison} Since it is not possible to know the true FCNs for EEG data, we consider a data-driven way of generating "ground-truth" graphs similar to \cite{navarro2022joint}. In particular, we generate two different types of "ground-truth" graphs using svGL as follows. 

In the first case, we learn a "ground-truth" graph structure for each subject separately using svGL. For each subject $s$, a data matrix $\mX^s \in \setR^{64\times m}$ is constructed from all 34 $\mX^{st}$ matrices, where $m=34\times 50$ and 34 is the number of trials. A "ground-truth" graph $G^s$ with edge density of $0.15$ is then learned from $\mX^s$ using \svgl. Next, for each subject, we construct a partial data matrix $\widehat{\mX}^s \in \setR^{64 \times m_\rho}$ from the first $\lfloor \rho \times 34 \rfloor$ $\mX^{sk}$ matrices, where $0<\rho<1$ and $m_\rho = \lfloor \rho \times 34 \rfloor \times 50$. Graphs are then learned using \mvglone{} and \svgl{} using $\widehat{\mX}^s$ following the same setup used for simulated data and then compared to "ground-truth" $G^s$'s using F1 score. 

The accuracy of recovery results are reported in Figure \ref{fig:eeg_gt_based_comparison}a. for different values of $\rho$. As can be seen, \mvglone{} has higher performance than \svgl{} for values of $\rho$ up to 0.3. When the number of graph signals to learn from is low, \mvglone performs better, indicating that the proposed method can share information across different graphs improving the accuracy. For $\rho>0.3$, the performance of \mvglone and \svgl{} become similar to each other. The figure also reports the performance of consensus graph learned by \mvglone{}, where the learned consensus graph is compared to each "ground-truth" $G^s$ and average F1 score across subjects is reported. Since the consensus graph only captures the shared structure across subjects, it cannot distinguish between individual differences. Thus, its performance is lower than that for the view graphs learned by \mvglone{} and \svgl{}.

In the second case, a single "ground-truth" graph across all subjects is found using svGL. In particular, we construct a data matrix $\mX\in \setR^{64 \times m}$ by concatenating 34 trial data matrices $\mX^{sk}$ from all subjects, where $m=34\times 50 \times 20$ and 20 is the total number of subjects. \svgl{} is then applied to $\mX$ to learn a graph $G$ with edge density of $0.15$. $G$ is considered to be the ground truth graph for all subjects. \mvglone{} and \svgl{} are then applied to previously constructed partial data matrices, i.e. $\widehat{\mX}^s$'s, to learn the subject graphs. F1 scores are reported in Figure \ref{fig:eeg_gt_based_comparison}b., where \mvglone{} is observed to have much higher performance than \svgl{} for all $\rho$ values. It is also observed that the consensus graph learned by \mvglone{} has the highest F1 value. This result is expected, since the "ground-truth" $G$ mostly includes shared edges across subjects as it is learned using signals from all subjects. Similarly, subject graphs learned by \mvglone{} are also better than \svgl{}, as they also capture the shared structure.
\begin{figure}[!b]
    \centering
    \includegraphics[width=0.8\columnwidth]{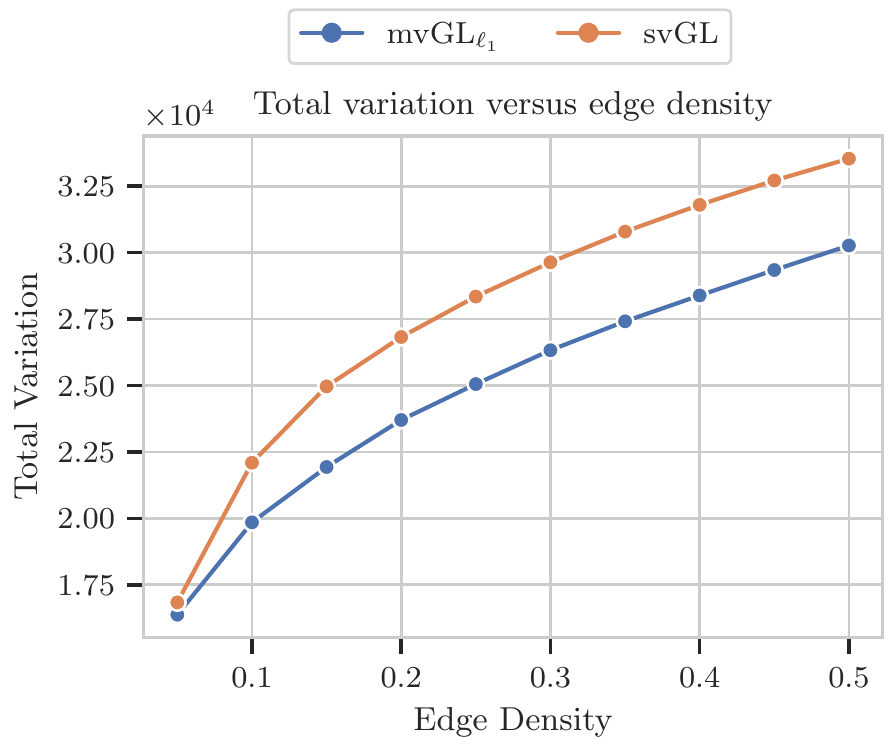}
    \caption{Total variation of test data on  subject graphs learned by different methods.}
    \label{fig:eeg_smoothness_based_comparison}
\end{figure}

These results indicate that the proposed method (i) provides better performance when the number of data samples is low by sharing information across views, (ii) can efficiently capture the shared structure across views, which in turn leads to better performance. 
\paragraph{Smoothness based comparison} Both \mvgl{} and \svgl{} rely on the assumption that the data is smooth with respect to the unknown graph structure. Therefore, we next compare their performances based on the smoothness of the learned graphs. To this end, we perform a five-fold cross validation by splitting 34 trials of each subject into five splits. One split is used as training set to learn the subject graphs, while the remaining ones are used as the test set, whose total variation (see \eqref{eq:smoothness}) with respect to the learned subject graphs is calculated. \svgl{} and \mvglone are used to learn the subject graphs with varying edge densities and $\beta$ parameter of \mvglone{} is set to a value where average pairwise correlation between subject graphs is around $0.75$. Average total variation of the test dataset is reported in Figure \ref{fig:eeg_smoothness_based_comparison}. It can be seen that subject graphs found by \mvglone{} have lower total variation, indicating that \mvglone{} manages to learn graphs that fit the smoothness assumption better.
\paragraph{Community Detection} An important goal of FCN analysis is to identify the functionally related subnetworks. Community detection can be used  to partition the FCNs to achieve this goal \cite{sporns2016modular}. To further validate the subject graphs found by \mvgl{}, we detect the community structure of the learned graphs and compare them to well-known networks in literature. \mvglone{} is employed to learn subject graphs $\{G^s\}_{s=1}^{20}$ and a consensus graph $G^c$ using data from all available trials. We set $\alpha$ and $\beta$ such that the edge density and average pairwise correlation are 0.15 and 0.75, respectively. \svgl{} is also used to learn subject graphs, which is represented by $\{G_{sv}^s\}_{s=1}^{20}$. $\alpha$ parameter for \svgl{} is set similarly to have $0.15$ edge density.
\begin{figure}[t]
    \centering
    \includegraphics[width=0.9\columnwidth]{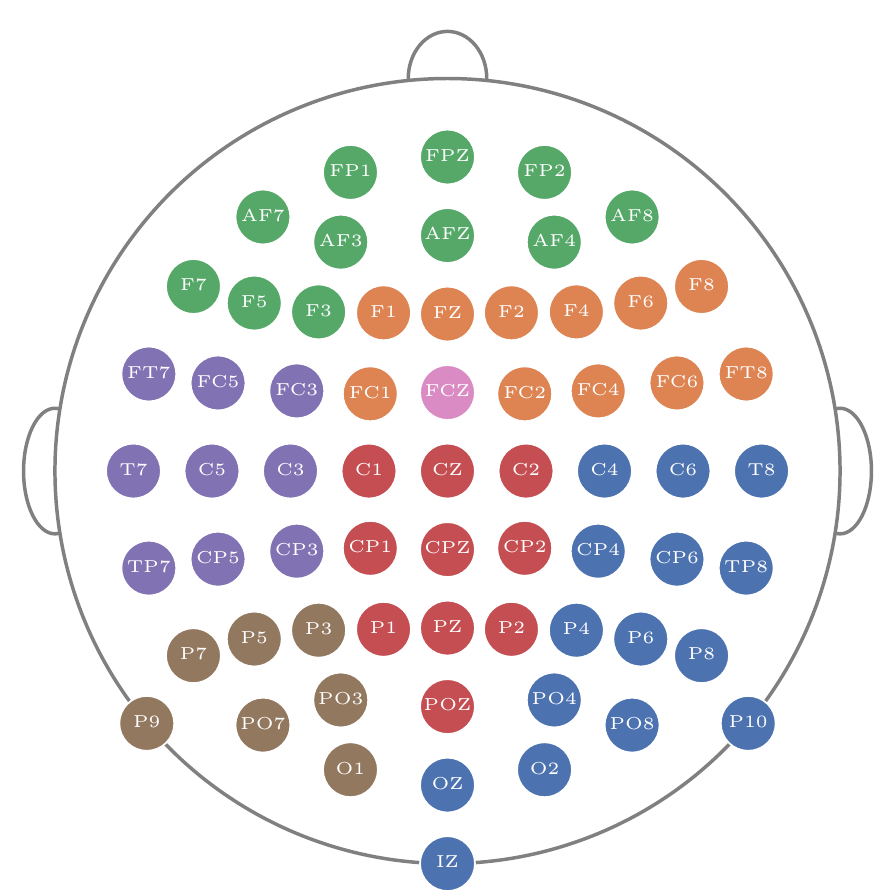}
    \caption{Community structure of the $G^c$ learned by \mvglone{}. Electrodes are colored based on their community memberships.}
    \label{fig:eeg_communities}
\end{figure}

One common approach for analyzing the community structure of FCNs across multiple subjects is to find the group community structure, that represents the shared partitioning across a group of subjects \cite{betzel2017multi}. Existing works usually perform this task by first finding each subject's community structure and then employing consensus clustering \cite{lancichinetti2012consensus}. Applying community detection to the consensus graph learned by the proposed method can eliminate this two-step process.

To this end, we found community structures of graphs learned by \mvglone{} and svGL by maximizing modularity \cite{newman2004finding}  using Leiden algorithm \cite{traag2019louvain}. Let $\{\calP^s\}_{s=1}^{20}$ be the subjects' community structure detected from $\{G^s\}_{s=1}^{20}$. Let $\calP^c$ be the community structure of $G^c$. Similarly, let $\{\calP_{sv}^s\}_{s=1}^{20}$ be the subjects' community structure found from graphs learned by svGL, i.e. $\{G_{sv}^s\}_{s=1}^{20}$. Finally, let $\calP_{sv}^c$ be the group community structure of $\{\calP_{sv}^s\}_{s=1}^{20}$ found by consensus clustering \cite{lancichinetti2012consensus}. We calculate how consistent consensus community structures, $\calP^c$ and $\calP_{sv}^c$, are with the subjects' community structures, $\{\calP^s\}_{s=1}^{20}$ and $\{\calP_{sv}^s\}_{s=1}^{20}$, using normalized mutual information (NMI) \cite{danon2005comparing}. The results are reported in Table \ref{tbl:consistency} and it is seen that $\calP^c$ is more consistent with both $\{\calP^s\}_{s=1}^{20}$ and $\{\calP_{sv}^s\}_{s=1}^{20}$. Thus, mvGL learns a consensus graph whose community structure is more consistent with the individual subjects' community structures.

We also study the community structure of $G^c$, shown in Figure \ref{fig:eeg_communities}, by comparing it to previous literature. The result indicates that there's a consensus community centered around frontal-central regions consistent with prior work indicating the increased activation of medial prefrontal cortex (mPFC) during cognitive control \cite{ozdemir2015hierarchical}. In addition, there are communities centered around the left and right lateral prefrontal cortices and the visual and motor regions.
\begin{table}[!t]
\centering
\caption{Consistency of group community structures with subjects' community structures as measured by NMI}
\label{tbl:consistency}
\begin{tabular}{p{0.31\columnwidth}p{0.27\columnwidth}p{0.27\columnwidth}}
      \toprule 
      & $\{\calP^s\}_{s=1}^{20}$ & $\{\calP_{sv}^s\}_{s=1}^{20}$ \\ \cmidrule{2-3}
      $\calP^c$ & 0.619 & 0.487 \\
      $\calP_{sv}^c$ & 0.418 & 0.427  \\
      \bottomrule
\end{tabular}
\end{table}
\section{Conclusions}
\label{sec:conclusions}
This paper introduced multiview graph learning based on the smoothness assumption for applications where multiple views of the same phenomenon are observed. The proposed framework learns both the individual view graphs as well as a consensus graph that captures the shared structured across views. The similarity across the different views is ensured through a consensus term between the individual view graphs and the consensus graph. The resulting optimization problem is formulated in a general way such that different functions could be employed for the consensus and regularization terms. The results illustrate the advantage of multiview graph learning over single graph learning when there's shared information across the views and when the data is noisy. 

Future work will explore the use of different functions for consensus and regularization such as information-theoretic functionals or generalized norms. While \mvglone{} and \mvgltwo{} focus on edge-based similarity across the views, they can also be extended to learn multiple graphs using the commonality of node-based structures, e.g. hub nodes \cite{mohan2014node}. The proposed approach's time complexity is quadratic in the number of nodes, which is not applicable to large-scale graphs. Future work will also consider scalability. 
\appendices
\section{Vectorization of \eqref{eq:multiview_graph_learning}}
\label{appdx:vectorization}
In order to show how to vectorize the optimization problem in \eqref{eq:multiview_graph_learning}, consider each term separately:
\begin{itemize}[leftmargin=*]
    \item For smoothness terms, let $\mK^i = \mX^i {\mX^i}^\top $, then we obtain $\trace({\mX^i}^\top \mL^i \mX^i) = \trace(\mX^i {\mX^i}^\top \mL^i) = \trace(\mK^i \mL^i)$, where the last term can be written as:
    \begin{align*}
        \trace(\mK^i \mL^i) 
            & = \sum_{a=1}^{n} \sum_{b=1}^{n} K_{ab}^i L_{ab}^i \\
            & = 2 \sum_{a=1}^n \sum_{b=a+1}^n K_{ab}^i L_{ab}^i 
                + \sum_{a=1}^n K_{aa}^i L_{aa}^i \\
            & = 2 {\vk^i}^\top \vell^i - {\vd^i}^\top \mS \vell^i = (2 {\vk^i} - \mS^\top \vd^i)^\top \vell^i,
    \end{align*}
    where in the second line we used the fact that both $\mK^i$ and $\mL^i$ are symmetric matrices and we used $\diag(\mL^i) = -\mS\vell^i$ in the third line.
    \item The Frobenius norm terms can be vectorized similarly where we use the symmetric structure of $\mL^i$ and $\diag(\mL^i) = -\mS\vell^i$:
    \begin{align*}
        \norm{\mL^i}_F^2  
            & = \sum_{a=1}^{n} \sum_{b=1}^{n} L_{ab}^i L_{ab}^i \\
            & = 2 \sum_{a=1}^n \sum_{b=a+1}^n L_{ab}^i L_{ab}^i 
                + \sum_{a=1}^n L_{aa}^i L_{aa}^i \\
            & = 2 {\vell^i}^\top \vell^i + {\vell^i}^\top \mS^\top \mS \vell^i = {\vell^i}^\top (\mS^\top \mS + 2\mI) \vell^i.
    \end{align*}
    \item Since all of the information for $\mL^i$ and $\mL$ is in the upper triangular parts of the matrices, $c(\cdot)$ and $r(\cdot)$ can simply be converted to $c_v(\cdot)$ and $r_v(\cdot)$, which return the same values given the upper triangular parts.
    \item For the constraint $\mL^i \in \setL$ (and $\mL \in \setL$), we use the definition of $\setL$ which includes $L_{ab}^i = L_{ba}^i \leq 0$ and $\mL \ones = 0$. The former implies $\upper(\mL^i) = \vell^i \leq 0$. $\mL\ones = 0$ is due to $\diag(\mL^i) = -\mS\vell^i$ and it can be ignored as we are only learning $\vell^i$. 
    \item Finally, $\trace(\mL^i) = 2n$ constrains the sum of node degrees to be $2n$ in the learned $\mL^i$. Since the sum of node degrees can also be calculated by $-2\ones^\top \vell^i$, we have the constraint $\ones^\top \vell^i = -n$ in \eqref{eq:multiview_graph_learning_vec}.
\end{itemize}
\section{Solution for ADMM steps}
\label{appdx:admm_steps}
In this section, the optimization of two ADMM steps (\eqref{eq:admm_first} and \eqref{eq:admm_second}) are given. As mentioned in the main text, \eqref{eq:admm_first} can be solved separately for $\vell^i$'s, $\vv^i$'s and $\vell$:
\begin{itemize}[leftmargin=*]
    \item \eqref{eq:admm_first} can be optimized for each $\vell^i$, separately. For this, rewrite the problem by ignoring all terms that do not depend on $\vell^i$:
    \begin{align}
        \label{eq:admm_li_step}
        \currval{\vell}^i & = \argmin_{\vell^i} f(\vell^i) + \imath_1(\vell^i)
                + {\prevval{\vy}^i}^\top(\prevval{\vz}^i - \vell^i)
                + \frac{\rho}{2} \norm{\prevval{\vz}^i - \vell^i}_2^2 \notag \\
            & = \argmin_{\vell^i}\ (2\vk^i - \mS^\top\vd^i)^\top \vell^i 
                    + \alpha {\vell^i}^\top(\mS^\top \mS + 2\mI)\vell^i \notag \\
                    & \phantom{= \argmin_{\vell^i}} + {\prevval{\vy}^i}^\top(\prevval{\vz}^i - \vell^i)
                    + \frac{\rho}{2} \norm{\prevval{\vz}^i - \vell^i}_2^2 \\ 
                & \phantom{=} \quad\ \textrm{s.t.} \quad\  \ones^\top \vell^i = -n, \notag
    \end{align} 
    where we substitute $f(\vell^i)$ and convert the indicator function to a constraint in the second equality. \eqref{eq:admm_li_step} is a quadratic problem with an equality constraint. Using its KKT conditions, its minimizer can be found as:
    \begin{align*}
        \currval{\vell}^i \hspace{-0.25em} = \hspace{-0.25em} \Pi_1[ 
            ( 2\alpha \mS^\top \mS + (4\alpha + \rho) \mI)^{-1}
           (\mS^\top \vd^i - 2\vk^i + \prevval{\vy}^i + \rho \prevval{\vz}^i)
        ]
    \end{align*}
    where $\Pi_1$ is the projection operator onto the hyperplane $\{\vell \in \setR^m | \ones^\top \vell = -n\}$.
    \item To solve \eqref{eq:admm_first} with respect to $\{\vv^i\}_{i=1}^N$, first define $\mV \in \setR^{m \times N}$ where columns of $\mV$ are $\vv^i$'s. Then, write \eqref{eq:admm_first} in terms of $\mV$ while ignoring all terms that do not depend on $\mV$:
    \begin{align}
        \currval{\mV} &= \argmin_{\mV}\ \beta c_v(\mV) + \frac{\rho}{2} \sum_{i=1}^N 
            \norm{\mV_{\cdot i} - \prevval{\vz}^i + \prevval{\vz} + \frac{1}{\rho}\prevval{\vw}}_2^2 \notag \\
            &= \argmin_{\mV}\ \beta c_v(\mV) + \frac{\rho}{2} \norm{\mV - \mA}_F^2, \label{eq:admm_vi_step}
    \end{align}
    where in the first step, we insert the scaled form of the augmented Lagrangian into \eqref{eq:admm_first}. In the second line, the summation term is written in a matrix form where $\mA \in \setR^{m \times N}$ with $\mA_{\cdot i} = \prevval{\vz}^i - \prevval{\vz} - 1/\rho \prevval{\vw}^i$. The optimization problem in \eqref{eq:admm_vi_step} is the proximal operator of $c_v(\cdot)$. For the proposed models in Section \ref{ssec:model_selection}, the proximal operator of $c_v(\cdot)$ has closed form solutions \cite{parikh2014proximal, bach2012optimization}.
    \item Finally, in order to solve \eqref{eq:admm_first} with respect to $\vell$; we rewrite it while ignoring the terms that do not depend on $\vell$:
    \begin{align}
        \currval{\vell} & = \argmin_{\vell} \gamma r_v(\vell) + \frac{\rho}{2} \norm{\prevval{\vz} - \vell + \frac{1}{\rho} \prevval{\vy}}_2^2, \label{eq:admm_l_step}
    \end{align}
    where we again use the scaled form of the augmented Lagrangian. \eqref{eq:admm_l_step} is the proximal operator of $r_v(\cdot)$. For \mvgltwo{}, $r_v(\cdot)$ is $\ell_1$-norm, whose proximal operator is the soft-thresholding operator \cite{parikh2014proximal}. 
\end{itemize}

The second step of ADMM given in \eqref{eq:admm_second} cannot be separated across its variables. However, we solve it with BCD where it is optimized with respect to $\vz^i$'s and $\vz$ alternatingly with the following subproblems:
\begin{align}
    \minimize_{\vz^i}\ & \imath_2(\vz^i) 
        + \frac{\rho}{2} \norm{\vz^i - \currval{\vell}^i + \frac{1}{\rho} \prevval{\vy}^i}_2^2 \notag \\
        & + \frac{\rho}{2} \norm{\currval{\vv}^i - \vz^i + \vz + \frac{1}{\rho} \prevval{\vw}^i}_2^2,\ \forall i \label{eq:admm_zi_step} \\
    \minimize_{\vz}\ & \imath_2(\vz) + \frac{\rho}{2} \sum_{i=1}^N \left\{
         \norm{\currval{\vv}^i - \vz^i + \vz + \frac{1}{\rho} \prevval{\vw}^i}_2^2 \right\} \notag \\
         & + \frac{\rho}{2} \norm{\vz - \currval{\vell} + \frac{1}{\rho} \prevval{\vy}}_2^2, \label{eq:admm_z_step}
\end{align}
which are derived from \eqref{eq:admm_second} by inserting the scaled version of augmented Lagrangian and ignoring the constant terms. At the $k$th iteration of BCD, $\vz$ in \eqref{eq:admm_zi_step} is set to the value obtained by solving \eqref{eq:admm_z_step} at the $(k-1)$th iteration. Similarly, $\vz^i$'s in \eqref{eq:admm_zi_step} are set to the values obtained by solving \eqref{eq:admm_zi_step} at the $k$th iteration. Both \eqref{eq:admm_zi_step} and \eqref{eq:admm_z_step} are proximal operators of $\imath_2(\cdot)$, which is equal to the projection onto $\setR_{\leq 0}^m$. 
%
\bibliographystyle{IEEEtran}
\bibliography{refs}

\end{document}